\documentclass[10pt,twocolumn]{article} 
\usepackage{simpleConference}
\usepackage{times}
\usepackage{graphicx}
\usepackage{amssymb}
\usepackage{url,hyperref}
\usepackage{subfigure}
\usepackage{booktabs}
\usepackage[authoryear,round]{natbib}
\usepackage{amsmath}
\usepackage{array}

\begin{document}
\bibliographystyle{plainnat}

\title{Definition-independent Formalization of Soundscapes: Towards a Formal Methodology}

\author{Mikel D. Jedrusiak$^{1}$, Thomas Harweg$^{1}$, Timo Haselhoff$^{2}$, Bryce T. Lawrence$^{3}$, Susanne Moebus$^{2}$, \\Frank Weichert$^{1}$ \\
\\
\small Department of Computer Science, Computergraphics, \\ \small Technical University of Dortmund, Otto-Hahn-Straße 16, 44227 Dortmund, Germany$^{1}$\\ \\
\small Institute for Urban Public Health, University Hospital Essen, University Duisburg-Essen, \\ \small Hufelandstraße 55, 45147 Essen, Germany$^{2}$ \\ \\
\small Department of Landscape Ecology and Landscape Planning, School of Spatial Planning, \\ \small Technical University of Dortmund, August-Schmidt-Straße 10, 44227 Dortmund, Germany$^{3}$ \\ \\ \\
\\
\\
}

\maketitle
\thispagestyle{empty}

\begin{abstract}
Soundscapes have been studied by researchers from various disciplines, each with different perspectives, goals, approaches, and terminologies. Accordingly, depending on the field, the concept of a soundscape's components changes, consequently changing the basic definition. This results in complicating interdisciplinary communication and comparison of results. Especially when soundscape-unrelated research areas are involved. For this reason, we present a potential formalization that is independent of the underlying soundscape definition, with the goal of being able to capture the heterogeneous structure of the data as well as the different ideologies in one model. In an exemplary analysis of frequency correlation matrices for land use type detection as an alternative to features like MFCCs, we show a practical application of our presented formalization.
\end{abstract}

\section{Introduction}\label{sec:introduction}

Soundscapes, which have gained in importance in recent years, are a central subject of various scientific research areas, such as bioacoustics, ecology, spatial planning or public health~\citep{DeCoensel:2010,Pijanowski:2011,Kang:2016,Kang:2018}. However, a precise definition of the term depends on the context of application or research, resulting in a variety of different conceptions. Based on~\citet{Southworth:1967} field study, in which he examined the characteristics of soundscapes in relation to people's perceptions, the understanding of the term soundscape has changed over the years depending on the field of research. A related definition was provided by~\citet{Schafer:1977} in his book ``The Tuning of the World'', in which he described soundscapes as the interaction of all acoustic events occurring in a single place. Since then, researchers from various fields of research have expanded and redefined the soundscape concept. For example, \citet{Pijanowski:2011}\ introduced the concept of soundscape ecology, which is dedicated to the study of how ``sounds can be used to understand coupled natural and human dynamics across different spatial and temporal scales''. \citet{Farina:2014}\ describe soundscapes as the counterpart to geographical and ecological landscapes, which share the spatial and time dimensions. In his book ``Soundscape Ecology'' various definitions are presented. Among others, soundscapes are described as an ``acoustic composition of voluntary and involuntary overlaps of different sounds with different origins and as an acoustic context that is produced and perceived in different ways by humans and animals''~\citep{Farina:2014Ecology}. In particular,~\citet{Pijanowski:2011} and~\citet{Farina:2014} categorize different sound sources into the categorizations Biophony, Geophony and Antrophony. In this context, Biophony includes all biological sound sources, such as the variety of distinctive animal sounds in a biome, whereas Geophony includes natural but non-biological sound sources~\citep{Krause:2008}. This involves sounds created by wind, water, weather, and geophysical forces, such as the acoustics of waves on a beach, the thunder in a thunderstorm, or an earthquake~\citep{Krause:2008}. Antrophony, by contrast, describes all sounds produced by humans. This includes speech, walking, music, traffic and many others~\citep{Pijanowski:2011}. In addition to the definitions mentioned here, several others exist. Accordingly, \citet{Barchiesi:2015} present soundscapes as equivalent to the acoustic scene and consequently as a composition of all sounds without consideration of human perception, while \citet{Westerkamp:2002} understands soundscapes as the result of the ``juxtaposition of environmental recordings that provide a sonic transmission of meanings about place, time, environment, and listening perception''. Furthermore, there also exists a definition from the International Organization for Standardization (ISO 12913-1:2014) that defines soundscapes as  ``acoustic environment as perceived, experienced and/or understood by one or more persons in context''~\citep{Iso:2014,Engel:2015}.\\
In order to identify the relationship between the different soundscape concepts, \citet{Grinfeder:2022} present an alternative partitioning into distal, proximal and perceptual soundscapes, restricting their focus to biophony and geophony, and excluding antrophony. In this context, distal soundscapes describe the spatial and temporal distribution of soundevents in a prespecified area and represent the total acoustic information. In the practical application, a specific time period as well as a specific acoustic range would be given.  Compared to the distal soundscapes, the proximal soundscapes depend on the potential position of the receiver, so they can be defined as the collection of propagated soundevents occurring at a specific point in space. Finally, the perceptual soundscapes express the individual subjective interpretation of the proximal soundscapes. Based on this, they present a visualization in their work that merges different soundscapes definitions.\\  
Our approach focuses on the concept of providing a formal view for the description of soundscapes, which considers not only acoustic information ($\mathcal{S}$) at different time points ($\mathcal{T}$), but also geospatial factors ($\mathcal{G}$), such as vegetation and built environment (e.g. distances to infrastructures such as main roads, parks, schools, etc.). We take into account that the human perception according to the ISO standard~\citep{Iso:2014} as well as the recording possibilities by sensor devices according to the soundscape ecology~\citep{Grinfeder:2022} are represented by our formalization ($\mathcal{A}$). In this way, we want to introduce $\mathcal{G}~\mathcal{A}~\mathcal{S}~\mathcal{T}$, intending to provide a definition-independent representation of soundscapes. The resulting advantages are a compact way to visualize the heterogeneous structure of soundscapes, as well as to improve interdisciplinary communication. Due to the definition independence, research approaches and results can be transferred to different research areas within the soundscape community without having to refer to the basic soundscape definition. Especially when research areas outside soundscape research are considered, an impact on interdisciplinary communication can be identified. In this paper, we present a classification example to demonstrate the application of our formalization by recognizing different land use types based on frequency correlation matrices~\citep{Haselhoff:2022}, which we calculate based on the Fourier transform of each day's recordings. At this point, we intentionally decide to use the frequency correlation matrices as an alternative feature to the classical ones, such as MFCCs~\citep{ittichaichareon2012speech}, in order to be able to study all the recordings of a day as a baseline for our analysis. Considering individual recordings of a location, correlations to the time of recording, day of the week, season or special events can be identified. In order to accurately capture this variety of a location, we need a significant number of recordings. Analyzing this amount of data in a coherent way also limits neural networks, because of limited GPU memory. FCMs provide the possibility of a compact representation of many single combined recordings to use them as input for network architectures.

In this context, we structure our work as follows. First, we summarize the current state of the art in Section~\ref{sec:stateOfTheArt}, focusing on current analysis techniques. In Section~\ref{sec:formalization}, we present our developed formalization, $\mathcal{G}~\mathcal{A}~\mathcal{S}~\mathcal{T}$. In the following Section~\ref{sec:analysis}, we describe the classification approach in~\ref{subsec:formalizationSALVE}, apply $\mathcal{G}~\mathcal{A}~\mathcal{S}~\mathcal{T}$ on the SALVE dataset and define our frequency correlation matrices in~\ref{subsec:fcm}. The methods used are detailed in Section~\ref{subsec:fcm_clustering} and~\ref{subsec:methods}, followed by a discussion of the results in Section~\ref{sec:discussion}. Finally, in Section~\ref{sec:conclusion} we give a brief outlook on our approach for future research.

\section{State of the Art}\label{sec:stateOfTheArt}

As indicated in Section~\ref{sec:introduction}, soundscapes are the focus of many different research areas, so different approaches to soundscape analysis exist. Primary research concentrates on various classification problems for determining the effect on humans using questionnaires~\citep{Sun:2019,Thorogood:2013}, and the recognition of individual sounds in recordings~\citep{Massoudi:2021}. Here, the analyses performed are based on mel-spectrograms~\citep{Arnault:2020,Massoudi:2021} as well as temporal features, spectral features, and cepstral features from signal theory (e.\,g., Zero Crossing Rate, Spectral Smoothness~\citep{oppermann2013reconstruction}, Mel Frequency Cepstral Coefficients (MFCC)\citep{ittichaichareon2012speech}, etc.)~\citep{Bountourakis:2015}. By using the determined quantities, the classification is performed by using statistical methods~\citep{Jeon:2015} or by applying machine learning techniques~\citep{Bountourakis:2015}. Among the statistical approaches, Principal Component Analysis (PCA) is used to extract the dominant factors, on which cluster analysis is followed using algorithms such as k-means~\citep{Jeon:2015,Sun:2019}. Another option is to use a Support Vector Machine (SVM), where in addition to the parameters already mentioned, acoustic descriptors, such as A-weighted energy-equivalent sound-pressure level, can be applied for classification~\citep{Torija:2014}. Another way to identify features is through unsupervised learning. \citet{Salamon:2015} use a two-stage architecture for this purpose, where the dimension of the data is reduced first with the help of a PCA, allowing a codebook to be learned based on this with a spherical k-means, and the result is then used by a classifier. A widely used possibility of classification based on learned features as well as on direct input of the mel spectrograms is the use of neural networks: The representation by spectrograms provides the possibility to use Convolutional Neural Networks~\citep{Massoudi:2021}. Especially the use of recurrent models like LSTMs~\citep{Lezhenin:2019} is reasonable because of the temporal dependencies in soundscapes. Combinations of both architectures are studied in the context of soundscape classification~\citep{Das:2020,Massoudi:2021}. 

\section{A formalization approach}\label{sec:formalization}
\begin{figure*}[t]
	\centering
	\includegraphics[width=0.8\textwidth]{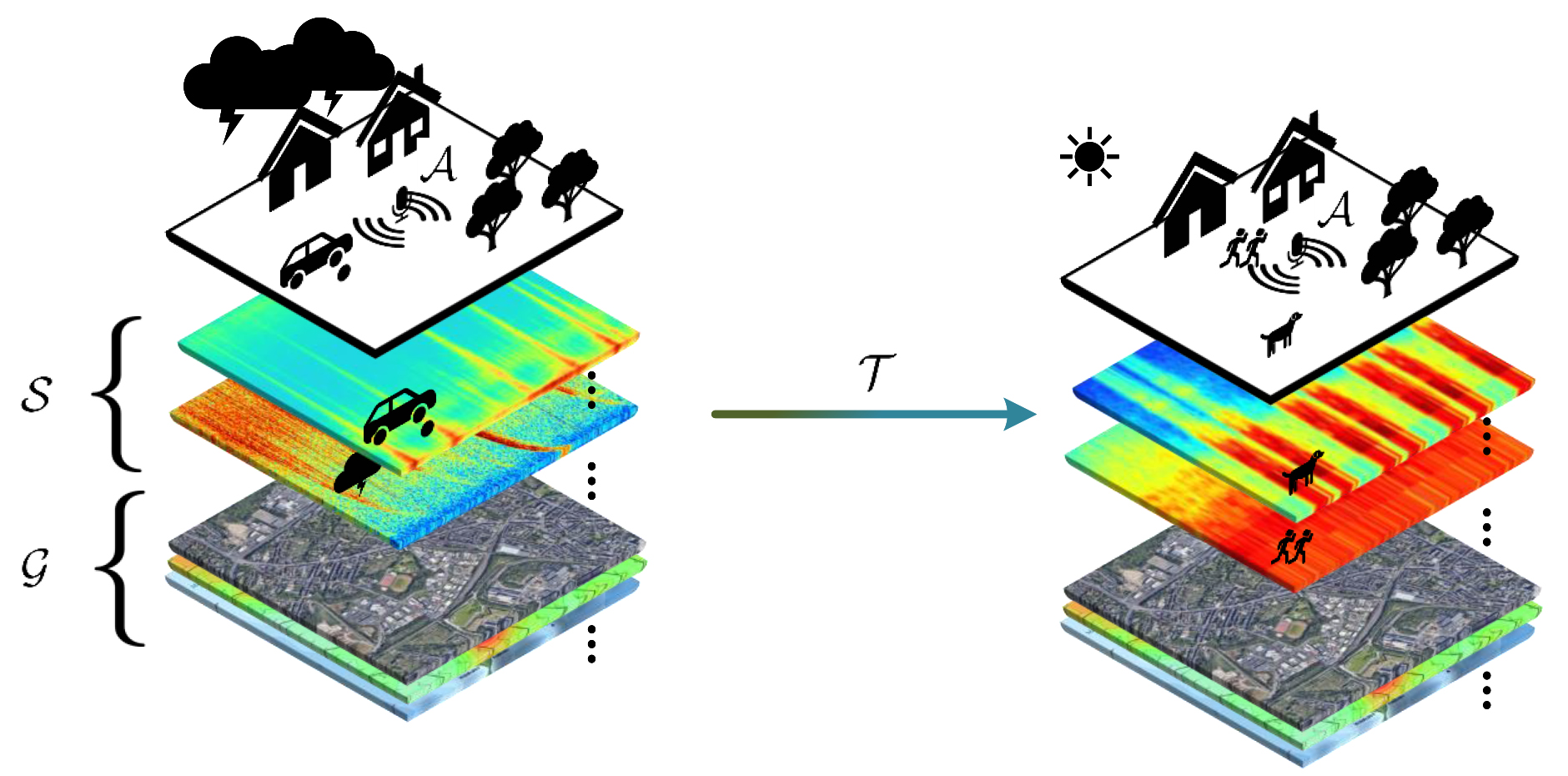} 
	\caption{The basic idea of our formalization of a time-dependent $\mathcal{T}$ layered model with sound sources $\mathcal{S}$ and geodata $\mathcal{G}$.}
\label{fig:layerModel}
\end{figure*}

Section~\ref{sec:introduction} indicates that there is no unique interpretation of soundscapes and that it varies depending on the research area. Consequently, there are challenges in both interdisciplinary communication and achieving comparability of analysis results. For this reason, we propose a formalization developed on the basis of existing definitions that can be used for the analysis of soundscapes. As already stated earlier, the proposed formalisation itself serves as a means for capturing the heterogenous structure of soundscapes, and only seeks to provide a common basis for all kinds of analyses. The methods for analyzing can be chosen independently afterwards.\\
Our concept of soundscapes is based on four components that capture the heterogeneous structure in the form of different partly time-dependent geographical data, as well as time-dependent audio data registered by a sensor. We define soundscape $\boldsymbol{x}$ by
\begin{align}\label{def:formalization}
\boldsymbol{x} = (\mathcal{G},\mathcal{A},\mathcal{S},\mathcal{T}),
\end{align}  
where $\mathcal{G}$ is the geographical component that impacts the soundscape, $\mathcal{A}$ represents the soundscape's audio signal, $\mathcal{S}$ includes the isolated signals from the physical sources specific to the given soundscape, and $\mathcal{T}$, which describes the temporal relationships. The audio signal represented by the set $\mathcal{A}$ corresponds to the overlapping of all occurring isolated signals perceived by a sensor or a human. The isolated signals in $\mathcal{S}$ can either be manually annotated or determined using sound event detection techniques.  In the same way, the influence of vegetation and infrastructure described by the set $\mathcal{G}$ affects the audio signal. Our general conception corresponds to a layer model, as visualized in Figure~\ref{fig:layerModel}, divided into a set of geolayers $\mathcal{G}$ defined by
\begin{align}\label{def:GeoLayer}
\mathcal{G}:=\{\mathcal{P}_{1},\dots,\mathcal{P}_{m_g}\}.
\end{align}
with $m_g\in\mathbb{N}$ and a set of source layers $\mathcal{S}$. Each layer represents a specific type of data required to characterize soundscapes. Typical geolayers $\mathcal{P}_i$ with $i\in\{1,\dots,m_g\}$ are for example point clouds, distance matrices to infrastructures, meteorological data, land use types, and others. The individual layers of $\mathcal{G}$ can be of different dimensionality, and additional time dependencies can occur, especially in the case of meteorological data. We model this time dependence with the set
\begin{align}\label{def:TimeLayer}
\mathcal{T} = \{\widetilde{\mathcal{T}_1},\dots,\widetilde{\mathcal{T}}_{m_t}\},
\end{align}
with $m_t\in\mathbb{N}$, whose elements correspond to an encoding of the temporal resolution of the recording. In this context, a $\widetilde{\mathcal{T}}_i$ with $i\in\{1,\dots, m_t\}$ can represent a time point as well as a temporal interval. In addition, for the elements of $\mathcal{T}$ an order is required. For example, the time points can be encoded by the year, month and day, so that an element of $\mathcal{T}$ is given by $\widetilde{\mathcal{T}}_i=(2022,10,5)\in\mathcal{T}$. The encoding may vary depending on the application scenario. The second type of layers are the source layers
\begin{align}\label{def:SourceLayer}
\mathcal{S}=\{\boldsymbol{s}_{\widetilde{\mathcal{T}}}\in\mathbb{R}^{m}\mid \widetilde{\mathcal{T}}\in\mathcal{T}\},
\end{align}
with dimension $m\in\mathbb{N}$, which contain isolated signals $\boldsymbol{s}_{\widetilde{\mathcal{T}}}$ from the physical sources with $\widetilde{\mathcal{T}}\in\mathcal{T}$ belonging to the soundscape. Examples of isolated signals include traffic sounds, people talking, animal sounds, and other soundevents that may occur as part of the environment similar to the definition of proximal soundscapes. 
The set $\mathcal{S}$ can e.\,g.\ be partitioned into the categories Biophony, Geophony, and Antrophony according to the definition of \citet{Pijanowski:2011}. By restricting $\mathcal{S}$ to signals that can be perceived and interpreted by people and by using human hearing as a sensor, we obtain the definition of ISO~\citep{Iso:2014} that a soundscape can be understood as the ``acoustic environment as perceived, experienced and/or understood by one or more persons in context''. As presented in Section~\ref{sec:introduction}, an elementary component of a soundscape is the specific interaction of the isolated signals in dependence of time. In our defined set 
\begin{align}\label{def:sensor}
\mathcal{A}:=\{\boldsymbol{o}\in\mathbb{R}^2,\phi\},
\end{align}
we specify the soundscape as recorded by a sensor or perceived by a human. This set contains the location coordinates $\boldsymbol{o}\in\mathbb{R}^2$ of the sensor as well as a sensor function
\begin{align}\label{def:RecordingFunction}
\phi := \sum_{\boldsymbol{s}\in\mathcal{S}}\omega(\boldsymbol{s})\oplus\sum_{\widetilde{\mathcal{T}}\in\mathcal{T}}\boldsymbol{r}_{\widetilde{\mathcal{T}}},   
\end{align}
which is defined on the basis of the isolated signals in $\mathcal{S}$ and a corresponding modulation function $\omega$.
By Formula~\ref{def:RecordingFunction}, $\boldsymbol{r}_{\widetilde{\mathcal{T}}}\in\mathbb{R}^{m}$ represents time-dependent noise. Two perspectives can be considered: On the one hand, we can interpret the formula as a recording by a sensor where we want to minimise the noise and the influence by the modulation function so that we can determine the isolated signals (manually or automatically). On the other hand, we can use the formula as a basis for creating soundscapes using already known isolated signals by determining a specific modulation function. An additional noise term can be added. Since $\phi\in\mathbb{R}^{m}$ we can use an index $k\in\mathbb{N}$ to get the $k$-th position in the vector with $\phi(k)\in\mathbb{R}$. In our studies, we initially assume that a soundscape is recorded by a single sensor. However, it is also possible to define a soundscape over multiple sensors by adding additional coordinates and sensor information to the set $\mathcal{A}$. This is especially important because several recording devices can record the same soundscape depending on their distance from each other. In a large forest area, sensors can record the same soundscape over several kilometers, while in a city different soundscapes can be recorded between two nearby areas such as a main street and a park. In general, our proposed formalization is independent of the point of view. This means, for example, that we can describe an urban soundscape with our formalization, which consists of different recording locations within the city. At the same time, we can differentiate the urban soundscape more by identifying each recording device as a soundscape, allowing us to separate, for example, main street and park.

\section{Classification of land use types}\label{sec:analysis}
\begin{figure*}[t]
\centering
\subfigure[Microphone location in a residential area.]{
\begin{minipage}[b]{0.43\linewidth}
	\includegraphics[height=5.25cm]{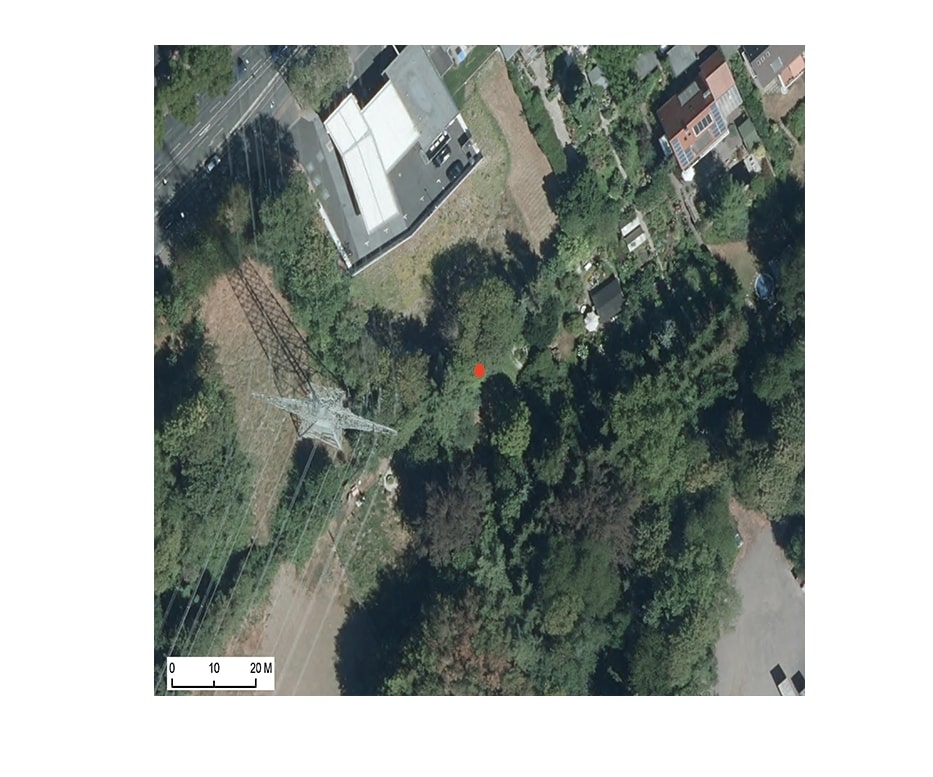}
\end{minipage}
\label{subfig:sat1}}
\subfigure[Microphone location in a residential street.]{
\begin{minipage}[b]{0.43\linewidth}
	\includegraphics[height=5.25cm]{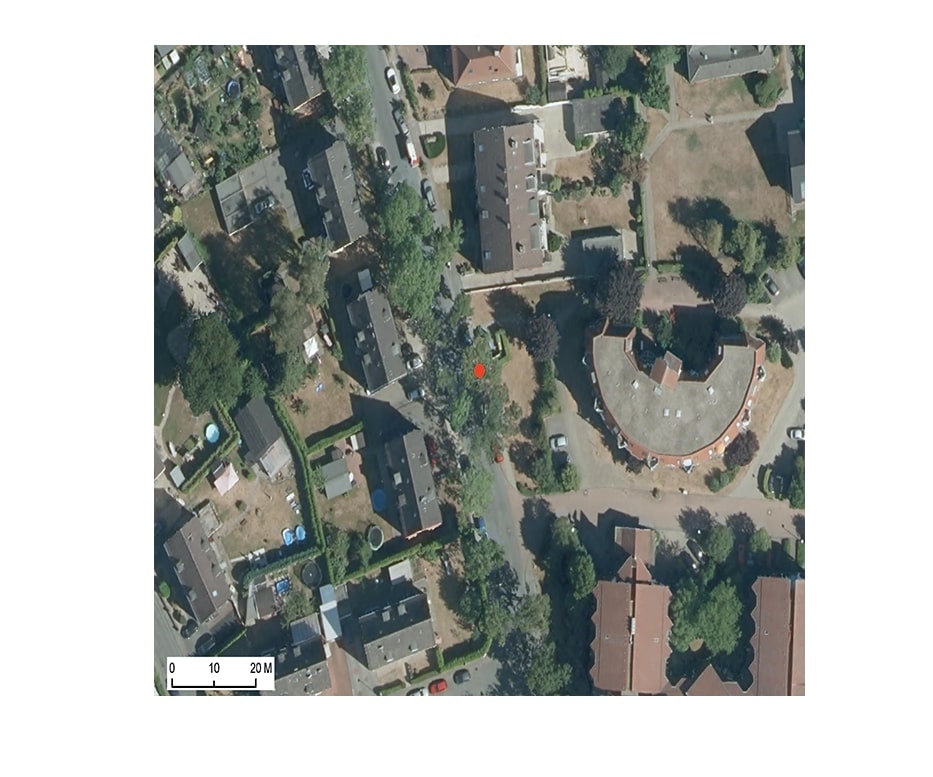}
\end{minipage}
\label{subfig:sat2}}
\subfigure[Frequency correlation matrix of a residential area.]{
\begin{minipage}[b]{0.43\linewidth}
	\includegraphics[height=5.25cm]{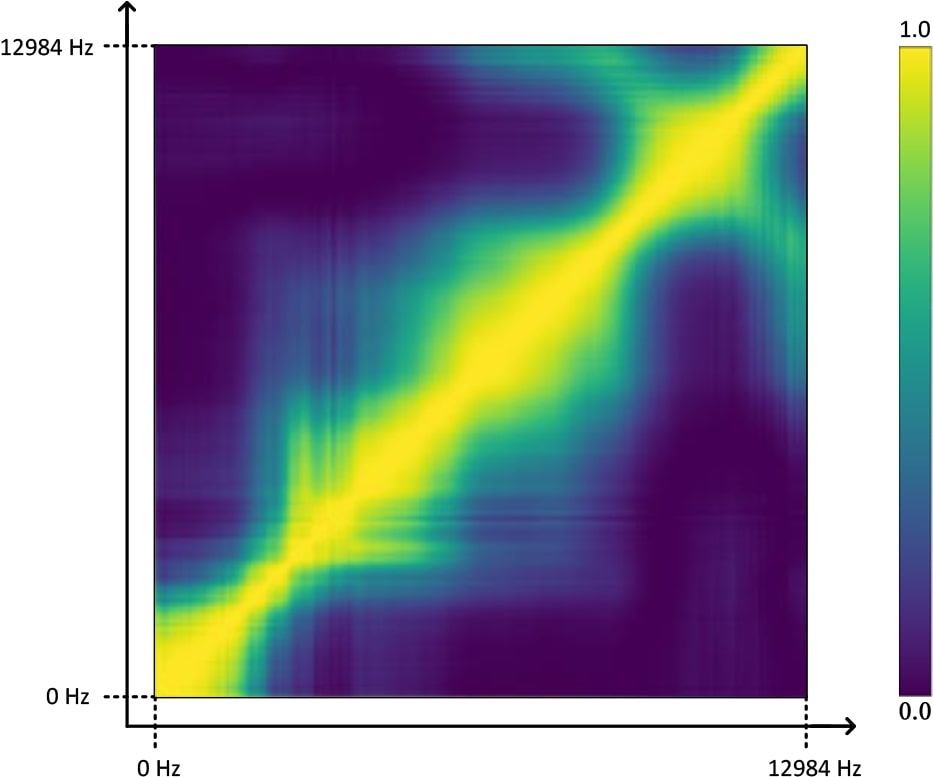}
\end{minipage}
\label{subfig:fcm1}}
\subfigure[Frequency correlation matrix of a residential street.]{
\begin{minipage}[b]{0.43\linewidth}
	\includegraphics[height=5.25cm]{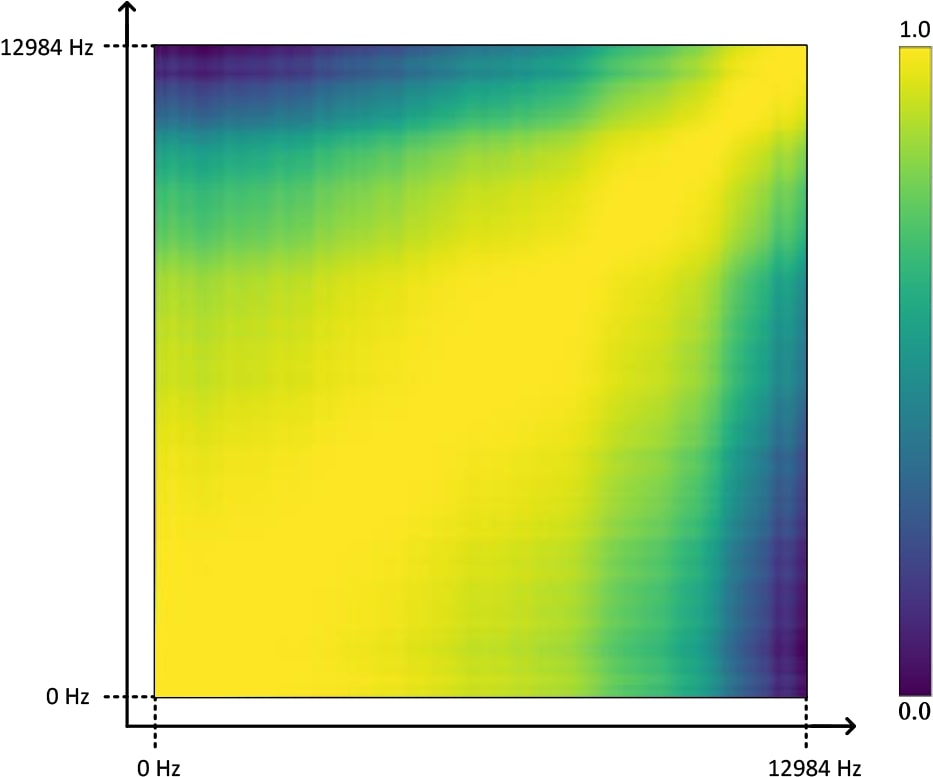}
\end{minipage}
\label{subfig:fcm2}}
\caption{Frequency correlation matrices based on all recordings of one day depending on their location. The scale used is the squared pearson coefficient.}\label{fig:fcm}
\end{figure*}
After introducing our formalization in Section~\ref{sec:formalization}, we use it to analyze soundscapes with respect to their localization. Our goal is to classify the given land use types considering the recordings made by the sensor $\mathcal{A}$ with respect to the time of recording $\mathcal{T}$. We use this exemplary analysis to demonstrate how to practically apply our formalization. 
\subsection{Formalization of SALVE data}\label{subsec:formalizationSALVE}
For our analysis, we use an already available data set from the SALVE study, defined in~\citet{Haselhoff:2022} as an automatic aural procedure (AAP\textsubscript{24}). This is a high temporal resolution sampling approach using stationary recorders. The dataset comprises 50 daily 3-minute recordings from 23 urban locations in Bochum, Germany. The Figures~\ref{subfig:sat1} and \ref{subfig:sat2} show examples of the recording locations. Recordings were made from May 2019 to the end of February 2020, resulting in 319,385 instances. The choice of the endpoint (February) is due to the COVID-19 pandemic and the resulting drastic changes in the acoustic environment after this date~\citep{Asensio:2020, Basu:2021, Hornberg:2021}. 
All recordings were made by using Wildlife Acoustics SM4 recorders with an SMX-SM4 Stub microphone~\citep{Acoustics:2020}. The Wildlife Acoustics SM4 Devices are not certified under any ISO. However, the recorders were calibrated according to Wildlife Acoustics and we recalibrated them as it was recommended after one year. The recorders clocks were synchronized in the beginning of the field phase. We used a MG 4010 Sound Calibrator with a pure 94~dB tone at 1~kHz and found that the 23 recorders were within the 3~dB variation that was marketed by Wildlife Acoustics. The devices were mounted at a height of approx.\ 1.65~m~\citep{DIN2018} and programmed to record 3-minute recordings every 26 minutes at a sampling frequency of 44.1~kHz and 16-bit depth. \\
According to our presented definition in Section~\ref{sec:formalization} we first define a quintuple composed of $t_1\in\{2019,2020\}$ for the year, $t_2\in\{1,\dots, 12\}$ for the month, and $t_3\in\{1,\dots,31\}$ for the day of the recordings. Since several recordings were made during one day, we specify the data by the number of hours $t_4\in\{0,\dots,23\}$, as well as the number of minutes $t_5\in\{0,20,40\}$. The time dependence is then given by $\widetilde{\mathcal{T}}_j=(t_1,t_2,t_3,t_4,t_5)\in\mathcal{T}$. Based on the structure of the elements of $\mathcal{T}$, it is intuitive to define a lexicographic order that represents the temporal relationship between the data. The coordinates $\boldsymbol{o}_i\in\mathbb{R}^2$ with $i\in\{1,\dots,23\}$ of the 23 different locations of the microphones used are given in the UTM coordinate system. We note the sensor information in the form $\mathcal{A}_i=\{\boldsymbol{o}_i,\phi\}$. We group our urban environments by land use. While the initial definition was based on the land use types (LUT) provided by the Ruhr Regional Association~\citep{Regionalverband:2020}, we defined new LUT categories that reflect the actual urban environment more appropriately, as devices were sometimes placed close to the edges of land use polygons. For this, we considered the original LUT, photographs, and assessments of the respective recording sites. Following, each team member did the classification separately. Disagreements were solved through discussion between the team members. Our used LUT categorizations are listed in Table~\ref{tab:confusion} distinguishing $n_\mathrm{LUT} = 9$ different land use types. To use our formalization, each soundscape is assigned an environment class $P_\mathrm{LUT}\in\{1,\dots,n_\mathrm{LUT}\}$. As an example, the class $P_\mathrm{LUT}$ = 1 represents a commercial area. The distribution of classes among recording locations is: one recorder for the playground class, two recorders each for the commercial area, green space, resedential area, urban agriculture, and urban forest classes, three recorders for the main street class, four recorders for the resedential street class, and five recorders for the small garden class. For the complete mapping between environment class and $P_\mathrm{LUT}$, we refer to Table~\ref{tab:confusion}. The geolayers of our soundscapes are given by $\mathcal{G}_i = \{P_\mathrm{LUT}\}$. In our particular case, isolated signals will not be analyzed in detail, so the set $\mathcal{S}$ can be left empty. In general, typical isolated signals in urban areas caused by people, cars and animals can be included in the set $\mathcal{S}$. Analogous to the Definition~\ref{def:formalization}, we represent our soundscapes in the form $\boldsymbol{x}_i=(\mathcal{G}_i,\mathcal{A}_i,\mathcal{S}_i,\mathcal{T})$.

\subsection{Frequency correlation matrices}\label{subsec:fcm}
For our studies on the relationship between the soundscapes and their recording location, we use frequency correlation matrices (\emph{FCM})~\citep{Haselhoff:2022a} based on the recordings of one day of one location. This allows us to study the impact of a weekday on the classification of land use types. At the same time, we reduce the amount of data from 50 recordings of one day to a frequency representation, so that the possibility of analysing several sequential days results. If we would work with classic features like MFCCs, we would already reach the limit of the used hardware when analyzing all recordings of one day, because especially a large amount of GPU memory is needed. For this reason, we are studying FCMs as an alternative feature, since it allows us to compactly represent several individual recordings. FCMs have been used in various research fields to study, e.g., cortical hubs in the human brain~\cite{Achard:2006}, but only seldom in the field of acoustics \cite{Nichols:2019}. In the context of urban acoustic environments (AEs), strong associations among frequency bands signify the prevalence of specific sound sources. This inter-bin correlations hold potential for distinguishing between AEs in different urban settings. To capture the frequency-based relationships that describe AEs, we employ Pearson's correlation coefficient to compute correlations between all frequency bins over time. In this context, $R^2$ values signify the linear connection between frequencies, which arises from sound sources closely following one another in time and spanning multiple frequency bins. Since the majority of sound source occurrences are mutually unrelated (e.g., birdsongs, sirens, etc.), this approach provides a means to differentiate various locations based on their distinct sound sources. However, due to the inherent overlap of diverse sounds within the urban acoustic environment, the primary purpose of FCMs is not to pinpoint individual sound sources, but rather to serve as a tool for characterizing the overall frequency dynamics of the specific urban acoustic environment over time. In the following, we describe how \emph{FCM} are calculated from the recordings of our dataset, with direct reference to our proposed formalization. In the first step, in order to obtain details about the frequency spectrum for each recording of a day, we calculate the spectrum with a fast Fourier transform (FFT). Afterwards they are merged to a matrix of frequency bands, where the two dimensions represent the number of frequency bands and the number of recordings. We express the number of recordings per day by $n_{r}$. Using the Fourier Transform $F(\cdot)$, for each recording we determine
\begin{align}
F_k(u) = \frac{1}{M}\sum_{j=0}^{M-1}\phi_k(j)\exp(-2\pi i\cdot\frac{ju}{M})
\end{align}
with recording $k\in\{1,\dots,n_{r}\}$, frequency $u\in\{1,\dots,M-1\}$ and $M\in\mathbb{N}$. In this way, we receive for each recording the frequency spectrum. Based on the recordings transformed into frequency domain, we subsequently determine the power spectrum $\boldsymbol{W}\in\mathbb{R}^{(M-1)\times n_{r}}$ with 
\begin{align}
w_{u,k} = |F_k(u)|.
\end{align} 
Subsequently, we arrange the resulting values into 1024 bins ($b\in\mathbb{N}$) of uniform size spanning from 0 to 22,050~Hz (with a bin width of 21.5~Hz and no spectral weighting). Within each bin, the values are energetically averaged~\cite{Haselhoff:2022a}
\begin{align}
\overline{w}_{i,j} = \frac{1}{b}\sum_{k=i\cdot b}^{(i+1)\cdot b-1}{w_{k,j}}.
\end{align}
To harmonize the magnitude of variability between these frequency ranges, a log-transformation is applied to all values. Consequently, we obtain the averaged power spectrum $\overline{\boldsymbol{W}}\in\mathbb{R}^{\lceil\frac{M-1}{b}\rceil\times n_{r}}$.
In order to enhance the robustness of our analyses, we employ Principal Component Analysis (PCA) to eliminate uncorrelated noise from individual frequency bins. The set of all FCMs with a selected frequency band is used as input for the PCA and is repeated with all frequency bands. In this process, we disentangle the temporal variability of each frequency bin into components that represent linear combinations of the initial data. These components are then sorted based on their eigenvalues, and we select the leading components for each frequency bin at every location, capturing $\geq 95~\%$ of the variance. By exclusively reconstructing each signal from these primary components--achieved through an inverse transformation--we derive de-noised time series data for each frequency bin. In the final step, we calculate the frequency correlation matrix $\boldsymbol{C}\in\mathbb{R}^{\lceil\frac{M-1}{b}\rceil\times\lceil\frac{M-1}{b}\rceil}$ using Pearson correlation $P(\cdot,\cdot)$ between all frequency bins $\overline{\boldsymbol{w}}_{i,\cdot}\in\mathbb{R}^{n_{\mathcal{M}}}$ and $\overline{\boldsymbol{w}}_{j,\cdot}\in\mathbb{R}^{n_{\mathcal{M}}}$
\begin{align}
c_{i,j} = P(\overline{\boldsymbol{w}}_{i,\cdot},\overline{\boldsymbol{w}}_{j,\cdot})
\end{align}
with $i,j\in\{1,\dots,M-1\}$, where $\overline{\boldsymbol{w}}_{i,\cdot}$ describes the $i$-th row of $\overline{\boldsymbol{W}}$. We report $R^2$ as a measure of the proportion of explained variance between two frequencies~\citep{waldmann2019use}. To characterize each day, we calculate \emph{FCM} for each date and recording device separately. In figures~\ref{subfig:fcm1} and~\ref{subfig:fcm2} are visualized examples of a residential area and residential street based on the quadratic correlation coefficient.
\subsection{FCM based clustering}\label{subsec:fcm_clustering}
\begin{figure}
\centering
\includegraphics[width=\linewidth]{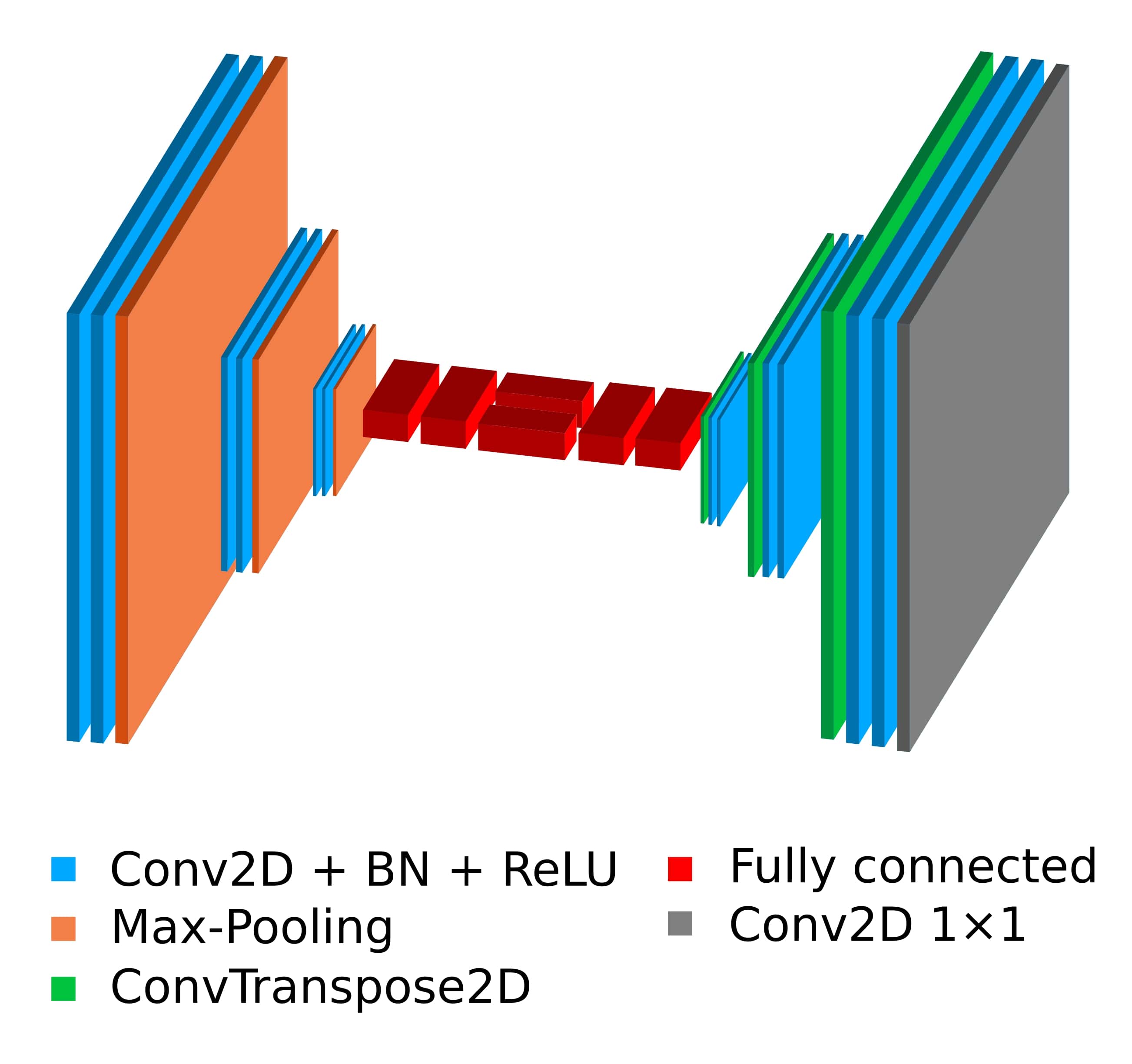}
\caption{Network architecture of the convolutional variational autoencoder (VAE). Single FCMs are used as input and target.}
\label{fig:conv_vae}
\end{figure}

\begin{figure*}[h!]
\centering
\subfigure[First cluster including the device ids 6, 10, 14 and 15]{
\begin{minipage}[b]{0.48\linewidth}
	\centering
	\includegraphics[height=6.2cm]{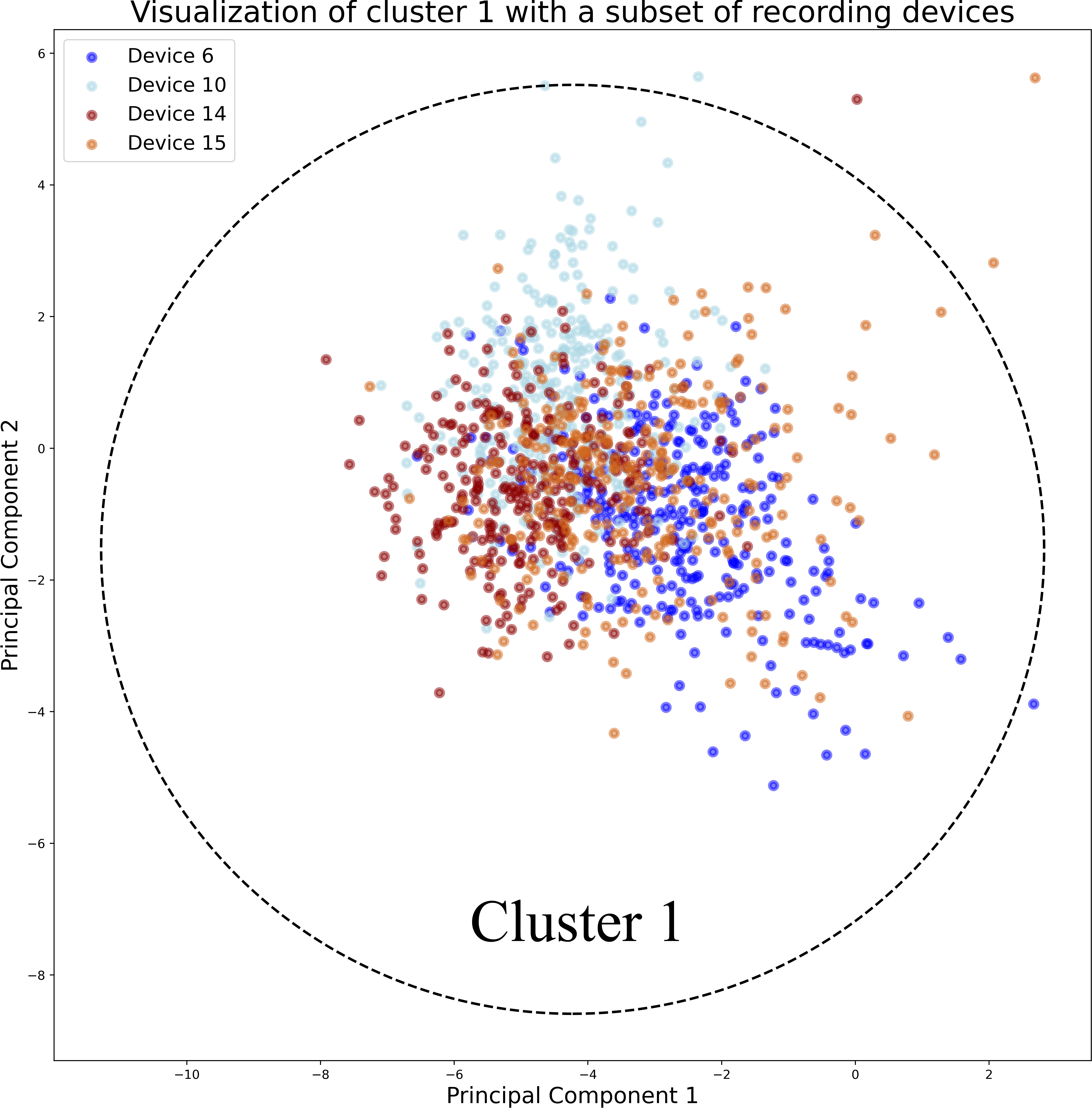}
\end{minipage}
\label{subfig:clust1}}
\subfigure[Second cluster including the device ids 2, 12, 20 and 24]{
\begin{minipage}[b]{0.48\linewidth}
	\centering
	\includegraphics[height=6.2cm]{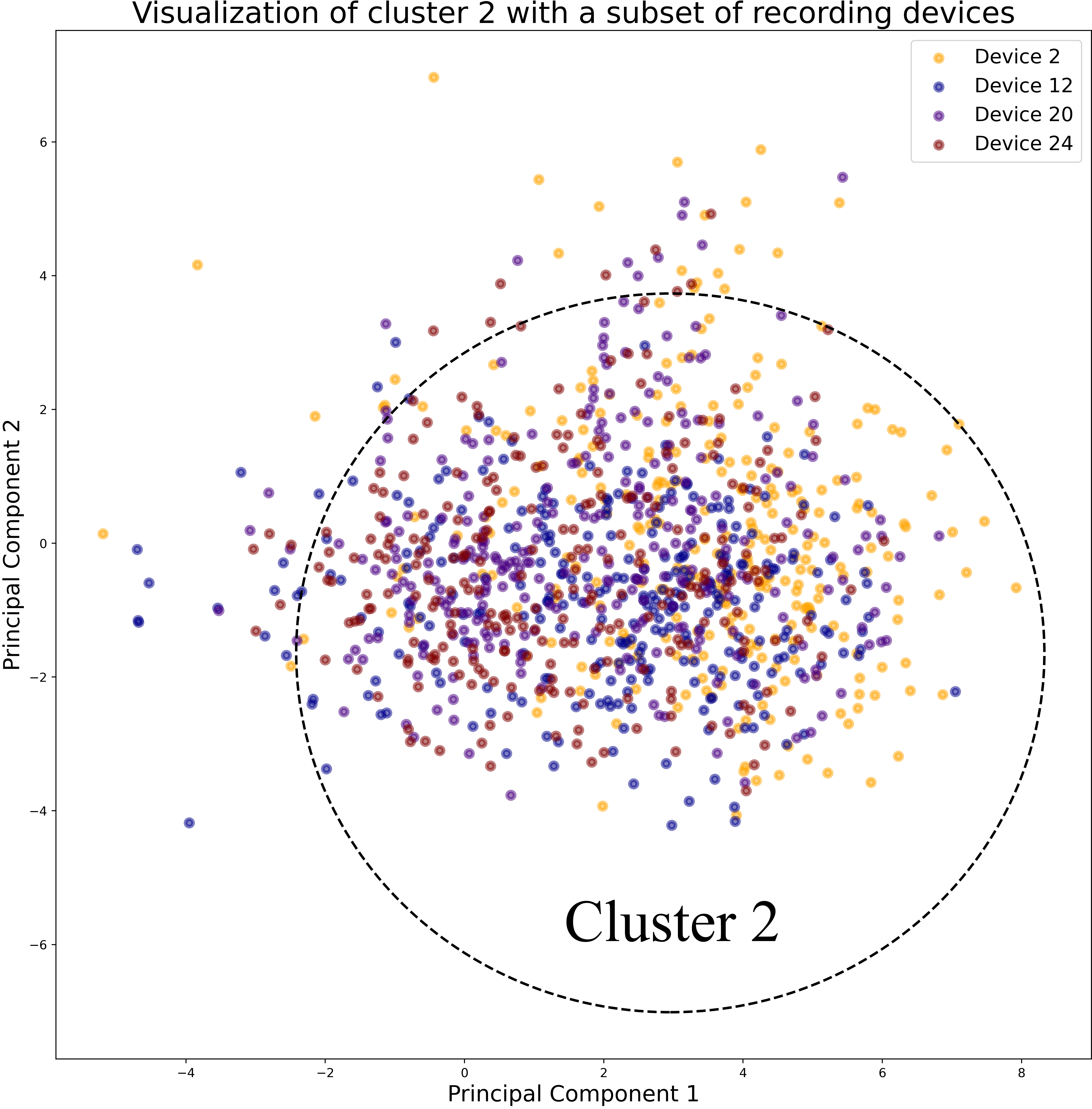}
\end{minipage}
\label{subfig:clust2}}
\subfigure[Third cluster including the device id 1]{
\begin{minipage}[b]{0.48\linewidth}
	\centering
	\includegraphics[height=6.2cm]{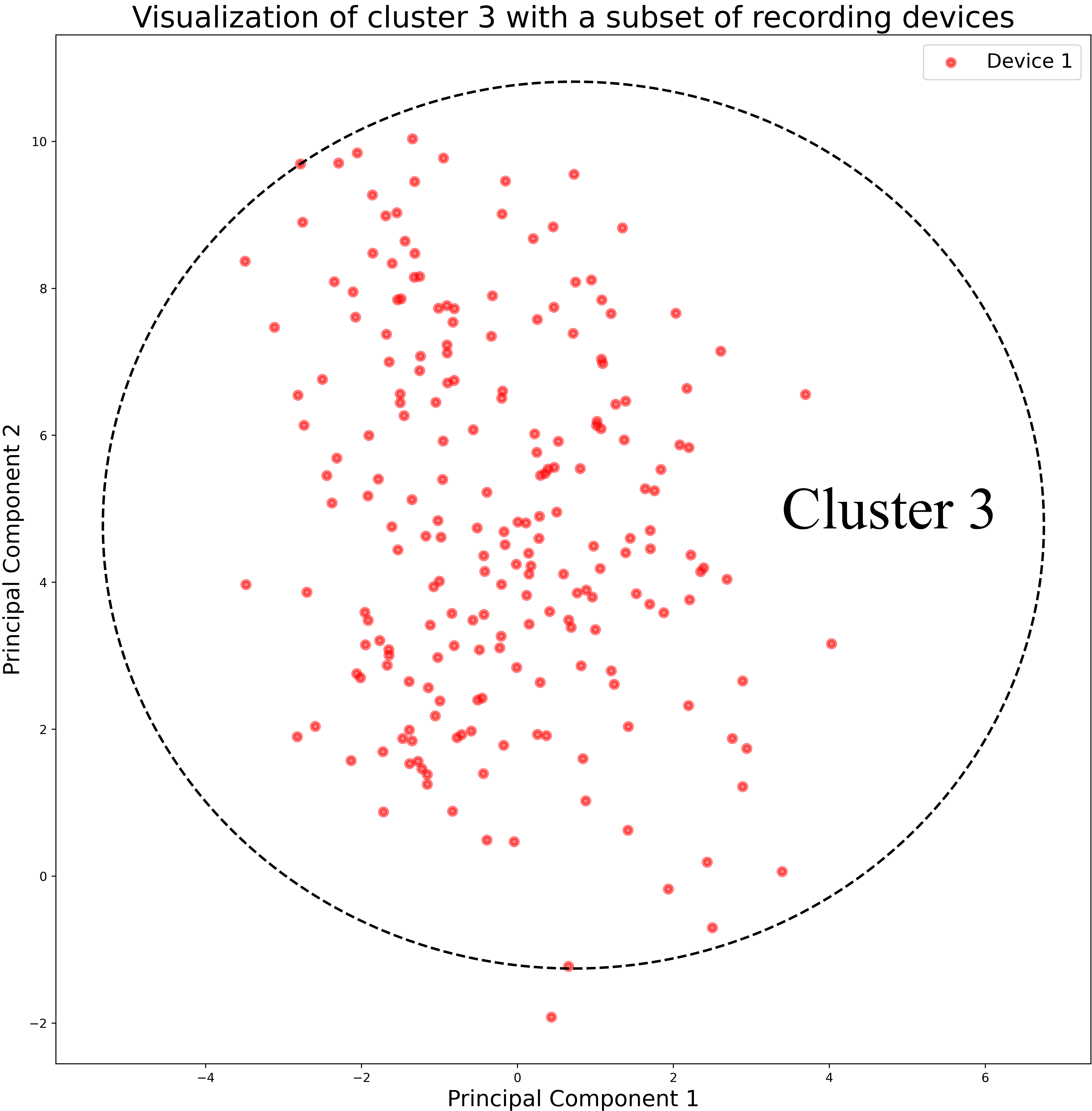}
\end{minipage}
\label{subfig:clust3}}
\caption{Our clustering result. Each point in a cluster corresponds to a \emph{FCM} encoded in the \emph{VAE}, with the color uniquely assigned to each recording device used.}\label{fig:clustering}
\end{figure*}

Based on the calculated FCMs, we first study the impact of an FCM on the classification of land use types by first considering each day of the week as equivalent. Generally, the day of the week, the season, or special events affect the FCM. In this first step of analysis, we first study the relationship between different locations and a frequency correlation matrix. For this purpose, we now identify feature vectors based on the FCMs, which we can cluster afterwards by applying k-means. Since on image data the latent space of a variational autoencoder (VAE) is commonly used as a feature vector for classification tasks~\citep{graving2020vae}, we apply this technique to our FCMs. 

As usual, the VAE can be divided into an encoder- and a decoder part, which are essentially symmetric. The encoder (or contracting path) consists of three convolutional sections, each of which consisting of two identical blocks, made up of a 2D-convolution with batch-norm and ReLU-activation each. The second of those blocks is complemented by a max-pooling layer at the end, for dimension reduction regarding the spatial extents. At the same time, the number of feature maps is doubled after each convolutional section. The output of the last convolutional block is then fed into a section of two fully connected layers, with the latter being
half the size of the former one. Afterwards, two additional fully connected layers map the result onto vectors $\boldsymbol{\mu}\in\mathbb{R}^{d_{\mathrm{l}}}$ and $\boldsymbol{\sigma}\in\mathbb{R}^{d_{\mathrm{l}}}$, representing mean and standard deviation of the variables constituting the latent space of dimension $d_{\mathrm{l}}$. These two layers operate on the same input and are on the same level, hierarchically. Contrary to the convolutional sections, we use PReLU-activations~\citep{He:2015} throughout the fully connected sections.

As noted, the decoder (or expanding path) of the VAE is constructed similarly. First, the latent space vectors are fed into a fully connected network with two layers, where the size is doubled from first to second. Then, an additional dense layer maps the output from the former to a two-dimensional shape of the same shape as the result of the last convolutional layer in the encoder section. From there on, three convolutional sections follow. These sections are similar to the encoder part, but starting with a transposed 2D-convolutional layer with batch-norm and ReLU, followed by a (regular) 2D-convolutional layer, again with batch-norm and ReLU activation. The number of feature maps is halved from one section to another in the expanding path. After the last convolutional layer of the decoder we add a $1\times 1$-convolutional layer to create the final output. The structure of the VAE is visualized in Figure~\ref{fig:conv_vae}. We train our VAE for 2000 Epochs with a learning rate of $10^{-5}$. Most dimensions of the VAE are determined by the size of the input image. The depth of the convolutional blocks, i.\,e.\ the number of feature maps, is 8, 16, and 32, respectively, for the encoder, and the same in reversed order for the decoder. We choose the size of the latent space as $d_{\mathrm{l}}=16$.\\

A common difficulty in training variational autoencoders is keeping the balance between reconstruction accuracy and adhering to the intended structure of the latent space~\citep{Bengio:2013, Higgins:2017}. To mitigate this issue, we employ the $\beta$-VAE disentangling strategy~\citep{Higgins:2017} for training, alongside an annealing weighting scheme~\citep{Bowman:2016}. With this strategy, the total loss $\mathcal{L}$ is calculated from the reconstruction loss $\mathcal{L}_{\mathrm{rec}}$ and the KL-divergence~\citep{Kingma:2014a, Kullback:1951}
loss $\mathcal{L}_{\mathrm{KL}}$ as $\mathcal{L} = \mathcal{L}_{\mathrm{rec}} + \beta \cdot \mathcal{L}_{\mathrm{KL}}$. The annealing scheme then determines the weighting factor $\beta$ depending on the current epoch. We use the linear scheme $\beta_e = \min\{e/e_{\mathrm{max}}, 1\} \cdot s$, with $e$ denoting the number of the current epoch, $e_{\mathrm{max}}=700$, and a constant scaling factor $s=10^{-4}$. We also employ a multi-scale loss strategy in order to force the VAE to preserve higher frequencies in the FCM images we feed into the model, as convolutional autoencoders tend to produce blurry results~\cite{Isola:2017, Pathak:2016}. The reconstruction loss $\mathcal{L}_{\mathrm{rec}}$ is comprised of a Laplacian pyramid loss~\citep{Bojanowski:2018} combined with the smooth-L1-norm~\citep{Girshick:2015}. To put more emphasis on higher frequencies, we also weigh the components of the Laplacian pyramid linearly with their level. This loss scheme seems sensible, as we train the VAE with rather small FCM images of size $128 \times 128$. 
The training is run on a subset of 100 randomly chosen FCM from the total set of 6389 recordings. The size of the subset was chosen empirically to cover the variation of FCM in the dataset, based on visual inspection and reconstruction accuracy. \\

Following the analysis of the VAE's latent space embeddings, we proceed to cluster the resulting vectors. To perform the clustering, we utilize the k-means clustering algorithm~\citep{Yang:2017}. However, before applying k-means, we first determine the optimal number of clusters by employing the elbow~\citep{Bholowalia:2014} and silhouette methods~\citep{Yuan:2019}. The elbow method involves plotting the explained variation as a function of the number of clusters and picking the elbow point in the plot. This point indicates where the addition of more clusters leads to a diminished improvement in the explained variation. On the other hand, the silhouette method measures the quality of clustering by calculating the average silhouette coefficient for each instance. This coefficient quantifies the similarity of an instance to its own cluster compared to other clusters. Considering both approaches, $n_c=3$ has been determined as the optimal number of clusters. Following the application of k-means algorithm, we finally obtain the clustering result presented in Figure~\ref{fig:clustering}. In Figures~\ref{subfig:clust1}--\subref{subfig:clust3}, the individual clusters are illustrated by dashed circles, with each point representing a dimensionally-reduced vector from the latent space, and thus corresponding to an \emph{FCM}. To calculate the cluster center, we average all points located in the cluster and determine the cluster circle based on this mean, which approximates the clusters for our visualization.  The color of each point is uniquely assigned to each recording device, and for a better overview we represent only a subset of the recording devices in the figure~\ref{fig:clustering}. 

Although the initial analysis helps identify distinct clusters, these clusters do not sufficiently correspond to the desired land use types. We can see that individual recording devices, with the exception of outliers, fall completely with their \emph{FCMs} into a cluster and are thus uniquely described by it, but in general the \emph{FCMs} of a device are distributed over several clusters.

\subsection{\emph{FCM} based classification network}\label{subsec:methods}
\begin{figure}[t]
\centering
\includegraphics[width=\linewidth]{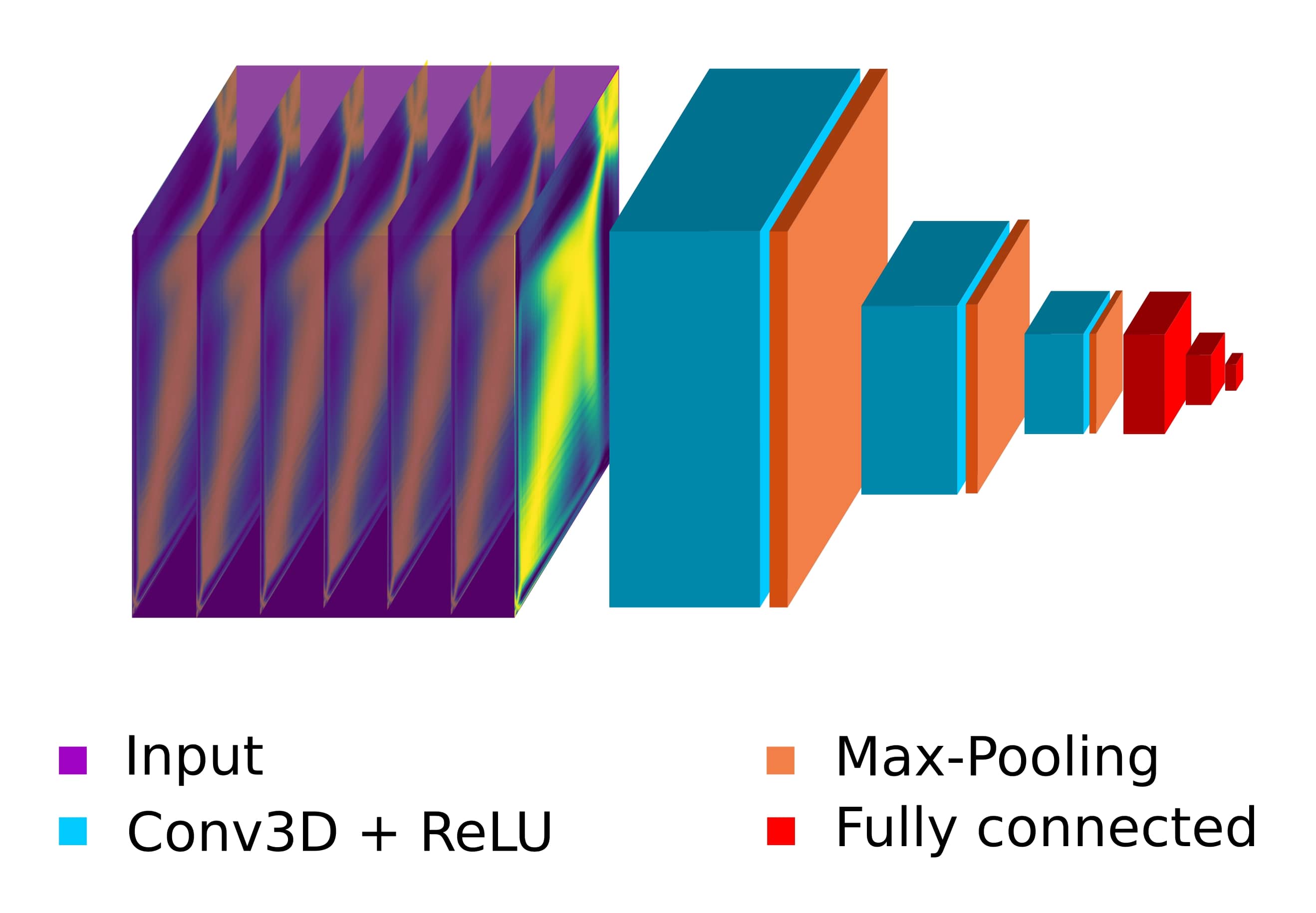}
\caption{Our architecture with a temporally sorted input of frequency correlation matrices for land use type classification.}\label{fig:envClassification}
\end{figure}
In our clustering analysis, we observed individual few recording devices that are already within a cluster with regard to their FCMs. These results can be explained by the influence of working days and weekends or special events.  For example, traffic on a road may be more intense on workdays than on weekends. Based on our results, we propose a time-based approach that takes as input a sequence of consecutive FCMs that are analyzed by applying 3D convolutional layers. To this end, we developed a network architecture as visualized in Figure~\ref{fig:envClassification}. As input for our architecture, we use FCMs of seven consecutive days with any day of the week as the starting point. By selecting seven days, we cover both weekdays and the weekend in one sequence, so that we cover the variation of a week. The dimension of the sequence as input to our network architecture is $7\times 256\times 256$. For feature extraction, we use iterative application of three-dimensional convolutional layers followed by maxpooling. We use a kernel $\boldsymbol{K}\in\mathbb{R}^{6\times 8\times 8}$ with 32 resulting output channels, which we shift with a step size of four along the x- and y-coordinates and with a step size of 1 along the z-axis. The choice of kernel size, especially along the time axis, is constrained by the maximum GPU memory available. 3D convolutional layers generate more feature maps compared to classic 2D convolutional layers, so significantly more memory is required. In our case, we used an RTX 2080Ti with 12 GB of memory, so the kernel size and output channels max out the memory of the graphics card. As an activation function we are using the ReLU-function. Kernel size and stride are defined analogously for the maxpooling layer. The results are then passed through fully connected layers, halving the number of neurons per layer until a layer with 64 neurons is reached. At this point, we use the softmax function as the activation with an output layer of nine neurons for each class.\\
Our dataset consists of the calculated \emph{FCMs} based on the daily recordings of 23 different devices at different locations. The groups are divided into disjoint sets of training, validation and evaluation data, i.e. a given sequence of seven days can lie exclusively in one of these sets. The sequences themselves, however, do not have to be disjoint. This means that in the evaluation set, there can be sequences that overlap with sequences in the training dataset. The identical sequence cannot occur. The decision to allow such overlaps is attributed to the dataset being used. If we were to insist on requiring disjunction among the sequences, the amount of data would be greatly reduced to the point where training would no longer be possible. We trained the architecture on 60~$\%$ of our dataset. Validation and evaluation took place on 20~$\%$ each. We trained our architecture on all environment classes, initially using data from one recorder per class. 

\begin{table*}[tb]
\caption{The table shows the results of different evaluation scenarios with precision (\emph{predicted positive value, PPV}), recall (\emph{true positive rate, TPR}) and F1 score.}\label{tab:confusion}
\centering
\begin{tabular}{l*{3}{>{\centering\arraybackslash}p{1.1cm}}*{6}{c}}
\toprule
& \multicolumn{3}{c}{Training and evaluation on} & \multicolumn{3}{c}{Evaluation on} & \multicolumn{3}{c}{Trainig and evaluation} \\ 
& \multicolumn{3}{c}{data from one device per class} & \multicolumn{3}{c}{unused devices} & \multicolumn{3}{c}{on all devices} \\
\cmidrule(r){2-4}\cmidrule(r){5-7}\cmidrule(r){8-10}
\textbf{$P_{\mathrm{LUT}}$. env. (devices)}	&\textbf{PPV} &\textbf{TPR} &\textbf{F1} &\textbf{PPV} &\textbf{TPR} &\textbf{F1}&\textbf{PPV} &\textbf{TPR} &\textbf{F1}\\
\midrule
1. com. area (2)	&1 &1 &1 &0.65 &0.05 &0.09 &1 &1 &1	 \\
2. green space (2) 	&1 &1 &1 &0.18 &0.08 &0.11 &1 &0.99&0.99 	 \\
3. main street (3)	&0.96 &1 &0.98 &0.62 &0.30 &0.40 &1 &1 &1	 \\
4. playground (1)  &1 &1 &1 &1 &1 &1 &1 &1 &1	 \\
5. res. area (2)	&1 &1 &1 &0.05 &0.03 &0.04 &1 &1 &1	 \\
6. res. streets (4)   &1 &1 &1 &0.32 &0.17 &0.24 &1 &1 &1	 \\
7. small garden (5)	&0.97 &0.98 &0.97 &0.3 &1 &0.46 &0.99 &1 &0.99	 \\
8. urban agriculture (2)	&1 &1 &1 &1 &1 &1 &1 &1 &1	 \\
9. urban forest (2)	&0.98 &0.94 &0.96 &0.28 &0.22 &0.25 &0.99 &0.99 &0.99	 \\	
\bottomrule
\end{tabular}
\end{table*}

\section{Discussion}\label{sec:discussion}
By analyzing FCMs in terms of classifying locations with given land use types, we have demonstrated how our formalization can be applied to a specific application. We have thus used our data in a definition-independent approach and, in the case of the FCMs, included our formalization in the computational method.  Our evaluation results can now be evaluated either purely methodologically or interpreted contextually with respect to the research domain. Another possibility is adaptation of the presented methodology to other inputs than FCMs, which basically require audio recordings. If a research area considers the ISO standard describing soundscapes as ``acoustic environment as perceived, experienced and/or understood by one or more persons in context'' as the underlying definition, a study can be performed analogously with the goal of determining the relationship between human perception and land use types. In the context of our formalization, the data collected in such a study can be described as derived data from $\phi$, allowing us again to achieve a definition-independent representation. If participants are exposed to only a single physical source, such as engine noise, the data can similarly be described as derived from the source layer $\mathcal{S}$. The resulting advantage is the comparability between the study that emphasizes human perception and the results obtained from the direct analysis of the data, allowing them to be discussed and evaluated in an interdisciplinary manner. For example, considering soundscape ecology following the categorization into distal, proximal, and perceptual soundscapes as proposed by~\citet{Grinfeder:2022}, it is now possible to examine the correlation between proximal and perceptual soundscapes concerning land use types, regardless of which research area and soundscape definition have contributed to the respective preliminary work.\\
Let us now consider the results of our classification of land use types, which we present in Table~\ref{tab:confusion}. In the first category, we evaluate on the same recording devices we used for training but on disjoint datasets, while in the second category, we base the evaluation on a dataset using all recording devices that were not included in the training. Significant differences between the results can be observed at this point, with a noticeable performance decrease when applied to new devices. Individual classes, such as playground and urban agriculture, can still be well differentiated from the other classes. In general, however, the precision and recall values are low for the other land use types. The commercial area and main street still have reasonably high precision in comparison with 65\% and 62\%, respectively, which corresponds to the probability that an according classified land use type actually corresponds to one of these two classes.  Additionally, the recall of the small garden is remarkable, which indicates that another land use type is never incorrectly classified as a small garden. This performance decrease can be explained by a lack of variation in the dataset. By using one recorder per land use type, we train on a balanced dataset while reducing the amount of data. In particular, it should be noted that land use types are not strictly separated areas, so locations of different types can be close to each other. For this reason, we extend our training process to include the previously unused device data and use the data split described in this section for training and evaluation. The results are shown in the third category in Table~\ref{tab:confusion}. Here, we observe a significant improvement in classification results per land use type. After 60~$\%$ of all seven day splits, based on the recordings of all used devices, are used for training, the evaluation data, consisting of 20~$\%$ of all seven day splits, can be clearly assigned to their respective land use types. This underlines the former assumption that environment classes contain a greater variety. However, precision, recall, and F1 being close to one shows that individual devices have a particular acoustic ``fingerprint'' that allows for an accurate prediction of their soundscape. At the time of the evaluation, data from additional locations in Bochum or another city were not available, so we cannot directly exclude an overfitting of the model. It is possible that instead of the general land use types a direct mapping to the individual locations has been learned. An indication for this could be the performance decrease during the evaluation on unknown recording devices. On the other hand, these results still show that FCMs can be used as features, especially when larger amounts of recordings are analyzed.\\
This way, we have made a purely methodical evaluation of our data. With our formalization, we want to support an interdisciplinary communication. Let us consider soundscapes as exclusively humanly perceptible according to the ISO definition.  On this basis, it is possible to study if humans can differentiate between land use types. In particular, the impact on humans can be explored and the correlation between them can be determined. In the case of soundscape ecology, the correlation between proximal and perceptual soundscape would be analyzed in this context. Our results and methodological approach can be considered independently of the research area, as our presented formalization does not require an underlying soundscape definition. Therefore, we provide a baseline that we continue to develop. With our current components time layer, geolayer and source layer we have captured the essential ones. But we do not exclude the possibility that additional components are needed to accurately and uniquely identify a soundscape so that any possible application from any research area can be described. Accordingly, it must be studied if additional layers need to be added to the formalization we have presented, or if the existing layers need to be differentiated more precisely to avoid limitations. 

\section{Conclusion and future work}\label{sec:conclusion}
In our work, we have introduced a formalization of soundscapes that allows the description of soundscapes for interdisciplinary collaboration regardless of the underlying definition, thus providing a common basis for analysis. Examples were used to show how existing definitions can be applied to our formalization without modification. In particular, we presented its application in the exemplary analysis of FCMs (which we investigated as an alternative feature for classification). Within this framework, we presented two methods based on FCMs for determining land use categories. In a first approach, we divided the FCMs projected in the latent space into classes using a clustering method. This resulted in three classes instead of the nine predefined LUTs. However, the investigation focused on individual FCMs corresponding to specific days of the week. In order to take into account the variations due to different weekdays in different locations, we used a 3D convolutional neural network together with an FCM sequence spanning seven consecutive days for classification. We found that basic classification was feasible, although a significant amount of training data was required. In general, we were able to demonstrate that FCMs can be used as an alternative feature to evaluate a large number of individual recordings. In further work we want to extend our formalization and investigate the heterogeneous structure of soundscapes in more detail. In doing so, we aim to find a comparison criterion for soundscapes on the basis of our formalization, so that soundscape classes can be precisely determined.  
\bibliography{lit.bib}

\end{document}